\documentclass[10pt,letter]{article}

\usepackage{amstext,amsgen,latexsym}
\usepackage{amstext,amssymb,amsfonts,latexsym}
\usepackage{theorem}
\usepackage{pifont}
\usepackage[dvips]{graphics,epsfig}

 \newcommand{\bs}{\bigskip}
 \newcommand{\ms}{\medskip}
 \newcommand{\n}{\noindent}
 \newcommand{\s}{\smallskip}
 \newcommand{\hs}[1]{\hspace*{ #1 mm}}
 \newcommand{\vs}[1]{\vspace*{ #1 mm}}

 \newcommand{\setempty}{\mathrm{\O}}

 \newcommand{\nat}{\mathbb{N}}
 
 \newcommand{\integer}{\mathbb{Z}}

 \newcommand{\co}{\mathrm{co}\mbox{-}}

 \newcommand{\ie}{\textit{i.e.},\hspace*{2mm}}
 \newcommand{\eg}{\textit{e.g.},\hspace*{2mm}}

 \newcommand{\CC}{{\cal C}}

 \newcommand{\reg}{\mathrm{REG}}
 \newcommand{\cfl}{\mathrm{CFL}}
 
 \def\bbox{\vrule height6pt width6pt depth1pt}

\theoremstyle{plain}
\theoremheaderfont{\bfseries}
 \newtheorem{theorem}{Theorem}[section]
 \newtheorem{lemma}[theorem]{Lemma}
 \newtheorem{proposition}[theorem]{Proposition}

 {\theorembodyfont{\rmfamily}
 }
 {\theorembodyfont{\rmfamily} \newtheorem{example}[theorem]{Example}}
 {\theorembodyfont{\rmfamily} }

 \newtheorem{claim}{Claim}

 \newenvironment{proof}{\par \noindent
            {\bf Proof. \hs{2}}}{\hfill$\Box$ \vspace*{3mm}}

 \newcommand{\qed}{\hfill$\Box$ \bs}

 \newcommand{\ceilings}[1]{\lceil #1 \rceil}
 \newcommand{\floors}[1]{\lfloor #1 \rfloor}

\newif\ifnotesw\noteswtrue
   {\ifnotesw\marginpar[\hfill\(\top\)]{\(\top\)}\fi}%
      {\ifnotesw\marginpar[\hfill\(\bot\)]{\(\bot\)}\fi}
      
\newcommand{\mnote}[1]%
   {\ifnotesw\marginpar%
	  [{\scriptsize\begin{minipage}[t]{\marginparwidth}
	  \raggedleft#1%
		  \end{minipage}}]%
	  {\scriptsize\begin{minipage}[t]{\marginparwidth}
	  \raggedright#1%
		  \end{minipage}}%
    \fi}

\newcommand{\ignore}[1]{}

\newcommand{\track}[2]{[{\tiny \begin{array}{c} #1 \\%
      #2 \end{array} }]}
\newcommand{\cent}{|\!\! \mathrm{c}}   
\newcommand{\dollar}{\$}
 
\begin{document}
\begin{center}
{\Large {\bf Swapping Lemmas for 
Regular and 
Context-Free 
Languages}} \bs\\
{\sc Tomoyuki Yamakami} \ms\\
{School of Computer Science and Engineering, University of Aizu} \\
{90 Kami-Iawase, Tsuruga, Ikki-machi, Aizu-Wakamatsu} \\
{Fukushima 965-8580, Japan} 
\end{center}

\pagestyle{plain}
\bs

\begin{quote}
\n{\bf Abstract.}
In formal language theory, one of the most fundamental tools, known as pumping lemmas, is extremely useful for regular and context-free languages. However, there are natural properties for which the pumping lemmas are of little use. One of such examples concerns a notion of advice,
which depends only on the size of an underlying input.
A standard pumping lemma encounters difficulty in proving 
that a given language is not regular in the presence of advice. We develop its substitution, called a swapping lemma for regular languages, to demonstrate the non-regularity of a target language with advice. For context-free languages, we also present a similar form of swapping lemma, which serves as a technical tool to show that certain languages are not context-free with advice.

\end{quote}

\section{Motivation and Overview}\label{sec:introduction}

Since their creation in the 1940's, formal languages and grammars have found many applications in, for instance, programming languages and artificial intelligence. 
A rapid development of the Internet, in particular, may treat one-way finite automata as a meaningful model of hand-held communication devices with a small amount of fast cache memory, which can process incoming information streamlined through a communication channel. 
Such types of automata are closely associated with the classical notions of {\em regular languages} and {\em context-free 
languages}, which were also classified respectively as 
type 3 and type 2 by Chomsky. These notions, which 
are foundations to formal language theory,   
have been a key subject of many core undergraduate 
curricula in computer science. The beauty of these notions comes from their simplicity and applicability. 

One of the most useful tools in formal language theory is so-called {\em pumping lemmas}  \cite{BPS61} for regular and context-free languages. Most undergraduate textbooks in this classic theory describe these lemmas (and their variants) as a powerful but fundamental tool. A standard methodology of proving the non-regularity of a target language, for instance, is to apply the pumping lemma to the language, deriving a desired contradiction if we begin with assuming the regularity of the language (see, \eg \cite{HU79} for its argument). For certain natural properties of languages, however, the pumping lemmas are not in the most useful form and we thus need to develop another form of lemmas to prove those properties. A typical example concerns ``advised computation.''

One way of enhancing the power of machine's language recognition is to provide a piece of {\em advice}, which is a string depending only on the length of input strings, to the original input strings. Karp and Lipton \cite{KL82} formulated a mechanism of such advice for polynomial time-bounded computation. Inspired by the work of Karp and Lipton, Damm and Holzer \cite{DH95} and Tadaki, Yamakami, and 
Lin \cite{TYL04} discussed the computational power of finite automata when advice is provided as supplemental information to their single read-only input tapes. See Section \ref{sec:notation} for the formal definition of automata that take advice. As in \cite{TYL04}, we use the notation $\mathrm{REG}/n$ to denote the family of languages recognized by deterministic finite automata if  advice of size exactly $n$ is provided {\em in parallel} to an input of size $n$. Similarly, we define $\mathrm{CFL}/n$ for the language family characterized by nondeterministic pushdown automata with such advice.

As is known, the advised family $\reg/n$ is quite different from $\reg$, the family of regular languages.
Typical context-sensitive languages, such as 
$L_{3eq} = \{a^nb^nc^n \mid n\geq0\}$, naturally fall into this advised family  $\mathrm{REG}/n$. On the contrary, certain deterministic context-free languages, such as $Equal = \{w\in\{0,1\}^*\mid \#_0(w)=\#_1(w)\}$ and $GT = \{w\in\{0,1\}^*\mid \#_0(w)>\#_1(w)\}$, where $\#_a(w)$ denotes the number of $a$'s in $w$, do not belong to $\mathrm{REG}/n$. 
{\em How can we prove this fact?} 
Now, let us try to prove that, for example,  $Equal$ is not in $\mathrm{REG}/n$ using the pumping lemma for regular languages. Assume otherwise that $Equal$ is characterized by a certain regular language $L$ with advice strings over an alphabet $\Gamma$. 
Now, we apply the pumping lemma by picking a pumping-lemma constant $m$ and then choosing an appropriate string $w$ 
(together with an advice string) in $L$ 
of length at least $m$. Using its decomposition $w=xyz$, the pumping lemma pumps this chosen string $w$ with advice given in parallel to $w$, generating a new series of strings of the 
form $xy^iz$. These pumped strings must belong to $L$; however, are they still in  $Equal$ with the appropriate advice strings? 
At this point, we encounter a serious flaw in our argument. 
The pumping process unwisely pumps the original string as well as the valid advice string.  
Since the pumping lemma forces the size of pumped strings to change, their associated pumped advice might no longer be ``valid'' advice for $Equal$. 
Therefore, we cannot conclude that the pumped strings are actually in $Equal$ with advice. To avoid this pitfall, we need to develop a new lemma, which keeps advice valid before and after the application of the lemma. 

In this paper, we shall present such a desired lemma, which we refer to as the {\em swapping lemma}, encompassing an essential nature of regular languages. In many cases, this new lemma is as powerful as the pumping lemma is. As  examples, we shall later  demonstrate, by a direct application of the swapping lemma,  that the context-free language $Pal = \{ww^{R} \mid w\in\{0,1\}^*\}$ (even-length palindromes), where $w^R$ is $w$ in reverse, and the aforementioned languages $Equal$ and $GT$ cannot be in $\mathrm{REG}/n$ (and therefore, they are not in $\mathrm{REG}$). The last two examples show a separation of $\mathrm{DCFL}$, the family of deterministic context-free languages, from the advised class $\mathrm{REG}/n$. This immediately yields the class separation 
$\mathrm{DCFL}/n\neq \mathrm{REG}/n$, which has not been known so far.  Our proof of the swapping lemma for regular languages is considerably simple and can be obtained from a direct application of the pigeonhole principle. 

Likewise, we also introduce a similar form of swapping lemma for context-free 
languages to deal with the non-membership to the advised family  $\mathrm{CFL}/n$. 
With help of this swapping lemma, as an example, we prove that the language $Dup=\{ww\mid w\in\{0,1\}^*\}$ (duplicating strings) is not in $\mathrm{CFL}/n$ (and therefore not in $\mathrm{CFL}$, the family of context-free languages). 
Another (slightly contrived) example is the language $Equal_{6}$ consisting of all strings $w$ over an alphabet of $6$ symbols together with a special separator $\#$ for which each symbol except $\#$ appears the same number of times in $w$. 
Since $Equal_{6}$ is in the complement of $\mathrm{CFL}$, denoted $\co\mathrm{CFL}$, we obtain a strong separation between $\mathrm{CFL}/n$ and $\co\mathrm{CFL}/n$; in other words,  
$\mathrm{CFL}/n$ is not closed under complementation. 
Our proof of the swapping lemma for context-free languages is quite different from a standard proof of the pumping lemma for context-free languages. Rather than using context-free grammars, our proof deals with a certain restricted form of nondeterministic pushdown automata to track down their behaviors in terms of transitions of first-in last-out stack contents. 

The main purposes of this paper are summarized as follows: (i) 
to introduce the two swapping lemmas for regular and context-free languages, (ii) to give their proofs by exploiting certain structural properties of finite automata, and (iii) to demonstrate the strong separations between $\mathrm{DCFL}/n$ and $\mathrm{REG}/n$ and between $\mathrm{CFL}/n$ and $\co\mathrm{CFL}/n$. 

We hope that the results of this paper should contribute to a  fundamental  progress of formal language theory and revive fresh interest in basic notions of regular languages and context-free languages.

\section{Notions and Notation}\label{sec:notation}

The {\em natural numbers} are nonnegative integers and we write $\nat$ to denote the set of all such numbers. For any two integers $m,n$ with $m\leq n$, the notation $[m,n]_{\integer}$ stands for the integer interval 
$\{m,m+1,m+2,\ldots,n\}$. 

An {\em alphabet} is a nonempty finite set and our alphabet is  denoted by either $\Sigma$ or $\Gamma$. A {\em string} over an alphabet $\Sigma$ is a series of symbols from $\Sigma$. In particular, the {\em empty string} is always denoted $\lambda$. 
The notation $\Sigma^*$ expresses the set of all strings over $\Sigma$. 
The {\em length} of a string $w$, denoted $|w|$, is the total number of symbols in $w$. For each length $n\in\nat$, we write $\Sigma^n$ (resp., $\Sigma^{\leq n}$) for the set of all strings over $\Sigma$ of length exactly $n$ (resp., at most $n$). For any non-empty string $w$ and any number $i\in[0,|w|]_{\integer}$, $pref_{i}(w)$ denotes the first $i$ symbols of $w$; namely, the substring $s$ of $w$ such that $|s|=i$ and $sx=w$ for a certain string $x$. In particular, $pref_{0}(w)=\lambda$ and $pref_{|w|}(w)=w$. 
Similarly, let $suf_{i}(w)$ be the last $i$ symbols of $w$. 
For any string $x$ of length $n$ and two arbitrary indices $i,j\in[1,n]_{\integer}$ with $i\leq j$, the notation $midd_{i,j}(x)$ denotes the string obtained from $x$ by deleting the first $i$ symbols and the last $n-j$ symbols of $x$; thus, $midd_{i,j}(x)$ contains exactly $j-i$ symbols taken from $x$. As a special case, $midd_{i,i}(x)$ expresses the $i$th symbol of $x$. It always holds that $x= pref_{i}(x)midd_{i,j}suf_{n-j}(x)$. 

For any language $L$ over $\Sigma$, the {\em complement} $\Sigma^* - L$ of $L$ is denoted $\overline{L}$ whenever $\Sigma$ is clear from the context. 
The {\em complement} of a family 
$\CC$ of languages is 
the collection of all languages whose complements are in $\CC$. 
We use the conventional notation $\co\CC$ to describe the complement of $\CC$. 

We denote by $\mathrm{REG}$ the family of {\em regular languages}. Similarly, the notation $\mathrm{CFL}$ represents the family of {\em context-free languages}. 
We also use the notation $\mathrm{DCFL}$ for the {\em deterministic 
context-free language} family. 
We assume the reader's familiarity with fundamental mechanisms of {\em one-tape one-head finite-state automata} and their variant, {\em pushdown automata}  
(see, \eg \cite{HU79} for their formal definitions). 
Concerning these machine models, $\reg$, $\cfl$, and $\mathrm{DCFL}$ can
be characterized by deterministic automata 
(or dfa's), nondeterministic pushdown automata (or npda's), and deterministic 
pushdown automata (or dpda's), respectively.

Now, let us briefly state the notion of {\em advice} in the form given in \cite{TYL04}, which is slightly different from \cite{DH95}.
First, we explain how to provide advice strings to finite automata using a ``track'' notation.  Consider a finite automaton with one scanning head moving on a read-only input tape, which consists of tape cells indexed with integers. For simplicity, the leftmost symbol of any input string is always placed in the cell indexed $1$. Now, we split the tape into two tracks. The upper track contains an original input $x$ given to the machine and the lower track carries a piece of advice, which is a string $w$ (over a possibly different alphabet) of the length $|x|$. 
More precisely, the tape contains $n$ tape cells consisting of the string $\track{x}{w} = \track{x_1}{\sigma_1}\track{x_2}{\sigma_2}\cdots \track{x_n}{\sigma_n}$, where $x=x_1x_2\cdots x_n$ and $w=\sigma_1\sigma_2\cdots\sigma_n$, in such a way that each $i$th cell contains the symbol $\track{x_i}{\sigma_i}$. The machine takes advantages of this advice $w$ 
to determine whether it accepts the input $x$ or not. 
To deal with all different lengths,  
advice is generally given as a form of function\footnote{In the original definition of advice functions by Karp and Lipton \cite{KL82}, these functions are not necessarily computable. This is mainly because we are interested in how much ``information'' with which each piece of advice provides an underlying machine, rather than how to generate such information.} $h$ (which is called an {\em advice function}) mapping $\nat$ to $\Gamma^*$, where $\Gamma$ is another alphabet, such that $|h(n)|=n$ for any length $n\in\nat$. 

The succinct notation $\reg/n$, given in \cite{TYL04}, denotes the collection of all languages $L$ over an alphabet $\Sigma$ such that there are an advice function $h$ and a dfa $M$ for which, 
for all strings $x\in\Sigma^*$, $x\in L$ iff $M$ accepts $\track{x}{h(|x|)}$. 
Since each dfa $M$ characterizes a certain regular language, say, $L'$, the above definition of $\reg/n$ can be made machine-independent by replacing the dfa $M$ by the regular language $L'$: let $L\in\reg/n$ if
there exist an advice function $h$ and another language $L'$ in $\reg$ for which, 
for all strings $x\in\Sigma^*$, $x\in L$ iff $\track{x}{h(|x|)}\in L'$. Similarly, we can define the advised families $\mathrm{CFL}/n$ and $\mathrm{DCFL}/n$ from $\mathrm{CFL}$ and $\mathrm{DCFL}$, respectively.

To help the reader grasp the concept of advice, we shall see a quick example of how to prepare such advice and use it to accept our target strings. Consider the context-sensitive language $L_{3eq} = \{a^nb^nc^n\mid n\in\nat\}$. It is obvious that $L_{3eq}$ is not a regular language. Let us claim that $L_{3eq}$  belongs to $\mathrm{REG}/n$. 

\begin{example}\label{3eq-advice}
Consider the non-regular language $L_{3eq} = \{a^nb^nc^n\mid n\in\nat\}$. It is easy to check that $L_{3eq}$  belongs to $\mathrm{REG}/n$ by choosing an advice function $h$ defined as $h(n)= a^{n/3}b^{n/3}c^{n/3}$ if $n\equiv0\;(\mathrm{mod}\;3)$ and $h(n)=0^{n}$ otherwise. 
How can we recognize this language with advice? 
Consider a dfa $M$ that behaves as follows. 
On input $\track{x}{h(|x|)}$ with advice $h(|x|)$, if 
$x=\lambda$, then accept the input immediately. Otherwise, check if the first bit of $h(n)$ is $a$. If so,  check if $x=h(|x|)$ by moving the tape head one by one to the right.  This is possible by scanning the upper and lower tracks at once at each cell. 
If the first bit of $h(n)$ is $c$ instead, reject the input. It is obvious that $M$ accepts $\track{x}{h(|x|)}$ iff $x\in L_{3eq}$. Therefore, $L_{3eq}$ belongs to $\mathrm{REG}/n$. 
\qed
\end{example}

Another example is a context-free language $Pal = \{ww^R\mid w\in\{0,1\}^*\}$ (even-length palindromes), where $w^R$ denotes $w$ in reverse.  It is well-known that $Pal$ is located outside $\mathrm{DCFL}$; however, as we show below, 
advice helps $Pal$ sit inside $\mathrm{DCFL}/n$.

\begin{example}\label{pal-advice}
The non-``deterministic context-free'' language $Pal$ belongs to $\mathrm{DCFL}/n$. This claim is shown as follows. It is well-known that the ``marked'' language $Pal_{\#} =\{w\#w^{R}\mid w\in\{0,1\}^*\}$, where $\#$ is a center marker, can be recognized by a certain dpda, say, $M$ (see, \eg \cite{HU79} for its proof). The center marker in $Pal_{\#}$ gives $M$ a cue to switch the dpda's inner mode at the time the dpda's head moves from $w$ to $w^R$. More precisely, the dpda $M$ stores the left substring $w$ in its stack and, upon its cue from the marker, $M$ checks whether this stack content matches the rest of the tape content. Since there is no center marker in $Pal$, we instead use an advice function $h$ to mark the boundary between $w$ and $w^{R}$ in $ww^{R}$. We define  $h(0)=\lambda$, $h(n) = 0^{n/2-1}101^{n/2-1}$ if $n$ is even with $n\geq 2$, and $h(n)=1^n$ if $n$ is odd. The first occurrence of $10$ in $h(n)$ in the even case, for instance, signals the time of transition from $w$ to $w^R$ in a similar way as the dpda $M$ does for $Pal_{\#}$. This advice places $Pal$ in $\mathrm{DCFL}/n$. 
\qed
\end{example}

\section{Swapping Lemma for Regular Languages}

Our goal of this section is to develop a new form of useful lemma, which can substitute the well-known {\em pumping lemma} \cite{BPS61} for 
regular languages, even in the presence of advice. 
We have seen the power of advice in Examples \ref{3eq-advice} and \ref{pal-advice}: advice helps dfa's recognize non-``regular''  languages and also helps dpda's recognize non-``deterministic context-free'' languages. 
When we want to show that a certain language $L$, such as $Equal$ and $Pal$, does not belong to $\mathrm{REG}/n$, a standard way (stated in many undergraduate textbooks) is an application of the pumping lemma for 
regular languages. 
A basic scheme of the standard pumping lemma (and many of its variants) states that, for any infinite regular language $L$ and any string $w$ in $L$, as far as its string size is large enough (at least a certain constant that depends only on $L$), we can always pump the string $w$ (by repeating a middle portion of $w$) while keeping its pumped string within the language $L$. Unfortunately, as discussed in Section \ref{sec:introduction}, the pumping lemma is not as useful as we hope it should be, when we wish to prove that certain languages are located outside the advised family  $\mathrm{REG}/n$. To achieve our goal, we need to develop a different type of lemma, which we call the {\em swapping lemma} for regular languages.

We begin with a simpler form of our swapping lemma. 

\begin{lemma}\label{swapping-lemma}{\rm [Swapping 
Lemma for Regular Languages]}\hs{1}
Let $L$ be any infinite regular language over an alphabet 
$\Sigma$ with $|\Sigma|\geq2$. 
There exists a positive integer $m$ (called a swapping-lemma constant) such that, for any integer $n\geq 1$ and any subset $S$ of $L\cap\Sigma^n$ of cardinality more than $m$, the following condition holds: for any integer $i\in[0,n]_{\integer}$, there exist two strings $x=x_1x_2$ and $y=y_1y_2$ in $S$ with $|x_1|=|y_1|=i$ and $|x_2|=|y_2|$ satisfying that (i) $x\neq y$, (ii) $y_1x_2\in L$, and (iii)  $x_1y_2\in L$.
\end{lemma}

\begin{proof}
We prove the lemma by a simple counting argument with use of the pigeonhole principle. Let $L$ be any infinite regular language over an alphabet $\Sigma$. Choose a dfa $M = (Q,\Sigma,\delta,q_0,F)$ that {\em recognizes} $L$, where $Q$ is a finite set of inner states, $\delta$ is a transition function, $q_0\in Q$ is the initial state, and $F\subseteq Q$ is a set of final states. We define our swapping-lemma constant $m$ as $|Q|$. Let $n$ be any integer at least $1$ and let $S$ be any subset of $L\cap\Sigma^n$ with $|S|> m$. Clearly, $|S|\geq 2$. Choose an arbitrary index $i\in[0,n]_{\integer}$. If either $i=0$ or $i=n$, then the lemma is trivially true (by choosing any two distinct strings $x,y$ in $S$). Henceforth, we assume that $n\geq 2$ and $1\leq i\leq n-1$. 
Consider internal states just after scanning the $i$th cell. 
Since $|S|>|Q|$, there are two distinct strings $x,y\in S$ for which $M$ enters the same internal state, say $q$, after reading the $i$th symbol of $x$ as well as $y$. Since the dfa cannot distinguish $pref_{i}(x)$ and $pref_{i}(y)$ after reading these substrings, $M$
should accept the swapped strings  $pref_{i}(x)suf_{n-i}(y)$ and $pref_{i}(y)suf_{n-i}(x)$. This completes the proof.
\end{proof}

Notice that, in the no-advice case, our swapping lemma can be used as a substitute for the pumping lemma when a target language $L$ is not ``slim'' enough (\ie $|L\cap\Sigma^n|>m$).
Let us demonstrate two simple examples of how to use our swapping lemma. 
The first example is the context-free language $Pal = \{ww^R\mid w\in\{0,1\}^*\}$.

\begin{example}\label{pal-case}
The context-free language $Pal$ is not in $\mathrm{REG}/n$ (and thus not in $\reg$). Assume that $Pal$ belongs to $\mathrm{REG}/n$ and apply the swapping lemma for regular languages. Since $Pal\in\mathrm{REG}/n$, there are a language $L\in\mathrm{REG}$ 
over an alphabet $\Sigma$ and an advice function $h$ such that, 
for every string $x\in\{0,1\}^*$, $\track{x}{h(|x|)}\in L$ 
iff $x\in Pal$. 
Take a swapping-lemma constant $m$ that satisfies the swapping lemma for $L$. 
Choose $n = 2m$ and $i= n/2$. 
Let a subset $S$ of $L\cap\Sigma^n$ be  
$S = \left\{\track{x}{h(n)}\in L \mid |x|=n\right\}$. 
Notice that $|S|\geq 2^{n/2}>m$. By the lemma, there are two distinct strings $x,y\in\{0,1\}^n$ 
that force $\track{x}{h(n)}$ and $\track{y}{h(n)}$ to fall into $S$. 
By letting $u_1 = pref_{i}(x)$ and $u_2 = pref_{i}(y)$, the strings $x$ and $y$ are written as $x = u_1u_1^{R}$ and $y = u_2u_2^{R}$. 
Now, let us consider the two swapped strings $u_1u_2^{R} = pref_{n/2}(x)suf_{n/2}(y)$ and $u_2u_1^{R} = pref_{n/2}(y)suf_{n/2}(x)$. 
These strings are clearly not of the form $ww^{R}$, and thus the swapped strings $\track{u_1u_2^R}{h(n)}$ and $\track{u_2u_1^R}{h(n)}$ cannot belong to $L$. This is a contradiction against the swapping lemma. Therefore, $Pal$ is not in $\mathrm{REG}/n$. 
\qed
\end{example}

The use of the subset $S$ in the swapping lemma, Lemma \ref{swapping-lemma}, is of great importance in dealing with the advised family $\reg/n$ because, for instance, $S$ in the above example cannot be defined as $S= L\cap\Sigma^n$ in order to lead to a desired contradiction. There are also cases that require more dexterous choices of $S$. One of those cases is 
the non-regular language 
$Equal = \{ w\in\{0,1\}^* \mid \#_0(w)=\#_1(w) \}$.

\begin{example}\label{equal-case}
The deterministic context-free language $Equal$ is not in $\mathrm{REG}/n$. This statement was first stated in \cite{TYL04}. Our purpose here is to apply our swapping lemma to reprove this known result. Assume that $Equal$ is in $\mathrm{REG}/n$. 
There are a regular language $L$ and an advice function $h$ such that, for every binary string $x$, $x\in Equal$ iff $\track{x}{h(|x|)}\in L$. Take a swapping-lemma constant $m$ and set $n = 2m$ 
as well as $i=n/2$. In this example, we cannot take a subset   
$S = \left\{\track{x}{h(n)}\in L \mid |x|=n\right\}$ as we have done 
in Example \ref{pal-case}; instead, we rather choose $n/2+1$ 
distinct strings $w_0,w_1,w_2,\ldots,w_{n/2}\in\{0,1\}^n$, where $w_k = 0^{k}1^{n/2-k}0^{n/2-k}1^{k}$ for each index  $k\in[0,n/2]_{\integer}$, 
and we then define $S=\left\{\track{w_0}{h(n)},\ldots,
\track{w_{n/2}}{h(n)}\right\}$.  
Clearly, the cardinality $|S|$ is more than $m$. The crucial point of the choice of $w_{k}$'s is explained as  $\#_{0}(pref_{n/2}(w_k)) =k$ for any number $k\in[0,n/2]_{\integer}$.  
The swapping lemma provides two distinct strings 
$x=w_j$ and $y=w_k$ ($j\neq k$) such that 
$\track{x}{h(n)},\track{y}{h(n)}\in S$ and 
$\track{u_1}{h(n)},\track{u_2}{h(n)}\in L$, where 
the swapped strings $u_1$ and $u_2$ are of the form   
$u_1= pref_{n/2}(x)suf_{n/2}(y)$ and 
$u_2 = pref_{n/2}(y)suf_{n/2}(x)$. It easily follows that  
\[
\#_0(u_1) = \#_0(pref_{n/2}(x)) + \#_0(suf_{n/2}(y)) = j + \frac{n}{2}-k \neq \frac{n}{2}
\]
since $j\neq k$. This contradicts the result that 
$\track{u_1}{h(n)}\in  L$. 
Therefore, $Equal$ cannot be in $\mathrm{REG}/n$. 
\qed
\end{example}

Next, we present a more general form of our swapping lemma. In Lemma \ref{swapping-lemma}, two strings $x$ and $y$ are both split into two blocks and one of these blocks is used for swapping. In the next lemma, in contrast, these strings are split into any  fixed number of blocks, one of which is actually used for swapping. This form
is useful when we want to show that, for instance, 
the non-regular language $GT = \{w\in\{0,1\}^*\mid \#_0(w)>\#_1(w)\}$ does not belong to $\reg/n$.

\begin{lemma}\label{2nd-swapping-lemma}{\rm [Swapping Lemma 
for Regular Languages]}\hs{1}
Let $L$ be any infinite regular language over an 
alphabet $\Sigma$ with $|\Sigma|\geq 2$. There is a positive number $m$ such that, for any number $n\geq 1$, any set $S\subseteq L\cap\Sigma^{n}$, and any series $(i_1,i_2,\ldots,i_k)\in([1,n]_{\integer})^{k}$ with $\sum_{j=1}^{k}i_j\leq n$ for a certain number $k\in[1,n]_{\integer}$, 
the following statement holds. If  $|S| > m$, then there exist two  strings $x = x_1x_2\cdots x_{k+1}$ and $y=y_1y_2\cdots y_{k+1}$ in $S$ with  
$|x_{k+1}|=|y_{k+1}|$  and $|x_{j'}|=|y_{j'}|=i_{j'}$ for each index $j'\in[1,k]_{\integer}$ 
such that, for every index $j\in[1,k]_{\integer}$, (i) $x\neq y$, (ii) 
$x_1\cdots x_{j-1}y_{j}x_{j+1}\cdots x_{k+1}\in L$, and (iii) $y_1\cdots y_{j-1}x_{j}y_{j+1}\cdots y_{k+1}\in L$. 
\end{lemma}

\begin{proof}
Note that, when $k=1$, this lemma is indeed Lemma \ref{swapping-lemma}. Consider a dfa $M=(Q,\Sigma,\delta,q_0,F)$ that recognizes $L$. Let $S\subseteq L\cap\Sigma^{n}$ satisfy $|S| > m$, where $m = |Q|^k$. To  each string $s\in S$, we assign a $k$-tuple $(q_1,q_2,\ldots,q_k)$ of internal states of $M$ such that, for each $j\in[1,k]_{\integer}$, $M$ enters state $q_j$ after scanning the $(\sum_{e=1}^{j}i_e)$th cell. There are at most $|Q|^k$ such tuples. Since $|S| > |Q|^k$, there are two distinct strings $x,y$ in $S$ such that they correspond to the same series of internal states, say $(q_1,q_2,\ldots,q_k)$. 
Write $x = x_1x_2\cdots x_{k+1}$ and $y=y_1y_2\cdots y_{k+1}$, where $|x_{k+1}|=|y_{k+1}|$ and $|x_{j'}|=|y_{j'}|=i_{j'}$ for 
every index $j'\in[1,k]_{\integer}$.
Notice that, for each index $j\in[1,k]_{\integer}$, $M$ enters the same internal state $q_{j}$ after scanning $x_{j}$ as well as  $y_{j}$ on the inputs $x$ and $y$, respectively. Fix an index $j\in[1,k]_{\integer}$ arbitrarily. {}From the choice of $x$ and $y$, we can swap the two blocks $x_{j}$ and $y_{j}$ in $x$ and $y$ without changing the acceptance condition of $M$. Therefore, the swapped strings $x_1\cdots x_{j-1}y_{j}x_{j+1}\cdots x_{k+1}$ and $y_1\cdots y_{j-1}x_{j}y_{j+1}\cdots y_{k+1}$ are both accepted by $M$. 
This gives the conclusion of the lemma. 
\end{proof}

\begin{figure}[ht]
\begin{center}
\includegraphics*[width=13cm]{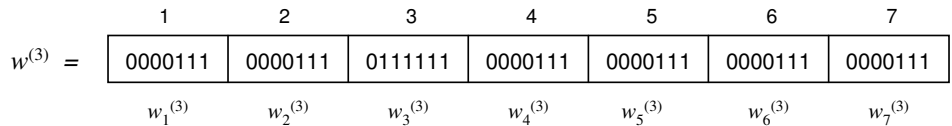}
\caption{A string $w^{(j)}$ with $j=3$ and $m=7$}\label{fig:GT-string} 
\end{center}
\end{figure}

Let us demonstrate how to apply Lemma \ref{2nd-swapping-lemma} to the deterministic context-free language $GT$.  

\begin{example}\label{GT-regular}
The deterministic context-free language $GT$ is not in $\reg/n$. Assuming that $GT\in \reg/n$, we choose an advice function $h$ and a regular language $L$ over an alphabet $\Sigma$ such that,
for every binary string $x$, $x\in GT$ iff $\track{x}{h(|x|)}\in L$. Since $L$ is an infinite regular language, we can apply Lemma \ref{2nd-swapping-lemma} to $L$. Let $m$ be a swapping-lemma constant. Without loss of generality, we can assume that $m$ is odd and at least $3$. Define $n = m^2$ 
and we are focused on the set $L\cap\Sigma^n$. 
Let $(i_1,i_2,\ldots,i_{m-1})$ be a unique series defined by $i_j=m$ for 
every index $j\in[1,m-1]_{\integer}$. This series makes each $n$ bit-string partitioned  into $m$ blocks of equal size $m$. For each index $j\in[1,m]_{\integer}$, let $w^{(j)}$ denote the string $w^{(j)}_1w^{(j)}_2\ldots w^{(j)}_m$ of the following form: (i) the $j$th block $w^{(j)}_j$ equals $01^{m-1}$ and (ii) any other block $w^{(j)}_i$ equals $0^{m'+1}1^{m'}$, where $m' = \floors{m/2}$. 
Figure \ref{fig:GT-string} gives an example of $w^{(j)}$ 
when $j=3$ and $m=7$.
Since $\#_0(w^{(j)}) = \#_1(w^{(j)})+1$, this string $w^{(j)}$ 
belongs to $GT$. 
The desired set $S$ is thus defined as  $\{\track{w^{(1)}}{h(n)},\ldots,\track{w^{(m)}}{h(n)}\}$. 
Clearly, $|S|\geq m$. 
By Lemma \ref{2nd-swapping-lemma}, there are two distinct 
strings $w^{(k)}$ and $w^{(l)}$ ($k\neq l$) such that $\track{w^{(k)}}{h(n)},\track{w^{(l)}}{h(n)}\in S$ and $\track{\tilde{w}^{(k)}}{h(n)}, \track{\tilde{w}^{(l)}}{h(n)}\in L$, 
where $\tilde{w}^{(k)}$ and $\tilde{w}^{(l)}$ are obtained respectively 
from $w^{(k)}$ and $w^{(l)}$ by swapping their $l$th blocks. 
Notice that, for each $i\in\{0,1\}$, 
$\#_i(\tilde{w}^{(k)}) = \#_i(w^{(k)}) - \#_i(w^{(k)}_{l}) + \#_i(w^{(l)}_{l})$.
Since $\#_0(w^{(k)}_{l}) = \#_1(w^{(k)}_{l})+1$, $\#_0(w^{(l)}_{l}) = 1$, 
and $\#_1(w^{(l)}_{l}) = m-1$, it immediately follows that 
$\#_{1}(\tilde{w}^{(k)}) = \#_{0}(\tilde{w}^{(k)}) + m-2$, 
which is greater than $\#_{0}(\tilde{w}^{(k)})$.  
This implies that $\tilde{w}^{(k)}\not\in GT$, 
contradicting the conclusion of Lemma \ref{2nd-swapping-lemma}. Therefore, $GT$ cannot be in $\reg/n$.
\qed
\end{example}

{}From Examples \ref{equal-case} and \ref{GT-regular}, since $Equal$ and $GT$ are both deterministic context-free, we can obtain a separation between $\mathrm{DCFL}$ and $\mathrm{REG}/n$. This gives a strong separation between $\mathrm{REG}/n$ and $\mathrm{DCFL}/n$, because  $\reg/n=\mathrm{DCFL}/n$ implies $\mathrm{DCFL}\subseteq \reg/n$ (using the fact that $\mathrm{DCFL}\subseteq \mathrm{DCFL}/n$).

\begin{proposition}
$\mathrm{DCFL}\nsubseteq \mathrm{REG}/n$. 
Or equivalently, $\mathrm{REG}/n \neq \mathrm{DCFL}/n$.
\end{proposition}

\section{Swapping Lemma for Context-Free Languages}\label{sec:CFL}

We have shown the usefulness of our swapping lemma for regular languages 
by proving that three typical languages cannot belong to $\mathrm{REG}/n$. Now, we turn our attention to $\mathrm{CFL}/n$, the family of context-free languages with advice. A standard\footnote{There are several well-known variants of pumping lemma for context-free languages. Ogden's lemma \cite{Ogd69}
is such a variant.} pumping lemma for context-free languages helps us pin down, for example, the language $Dup =\{ww \mid w\in\{0,1\}^*\}$ (duplicating strings) into the outside of $\mathrm{CFL}$. 
As in the case of regular languages, this pumping lemma is also of no use when we try to prove that $Dup$ is not in $\mathrm{CFL}/n$. 
This situation urges us to develop a new form of lemma, the {\em swapping lemma} for context-free languages. 

To state our swapping lemma, we introduce the following notation 
for each fixed subset $S$ of $\Sigma^n$. For any two indices 
$i,j\in[1,n]_{\integer}$ with $i+j\leq n$ and any string $u\in\Sigma^{j}$, the notation $S_{i,u}$ denotes the set $\{x\in S\mid u = midd_{i,i+j}(x)\}$. It thus follows that $S = \bigcup_{u\in\Sigma^j}S_{i,u}$ for each fixed index $j\in[1,n]_{\integer}$.
  
\begin{lemma}\label{swapping-lemma-CFL}{\rm 
[Swapping Lemma for Context-Free Languages]}\hs{1}
Let $L$ be any infinite context-free language over  an 
alphabet $\Sigma$ with $|\Sigma|\geq 2$. There is a positive number $m$ that satisfies the following. Let $n$ be any positive number at least $2$, 
let $S$ be any subset of 
$L\cap\Sigma^{n}$, and let  $j_0,k\in[2,n]_{\integer}$ be any two indices satisfying that $k \geq 2j_0$ and $|S_{i,u}|< |S|/m(k-j_0+1)(n-j_0+1)$ for any index $i\in[1,n-j_0]_{\integer}$ and any string $u\in\Sigma^{j_0}$. 
There exist two indices $i\in[1,n]_{\integer}$ and $j\in[j_0,k]_{\integer}$ with $i+j\leq n$ and two strings $x =x_1x_2x_3$ and $y=y_1y_2y_3$ in $S$ with $|x_1|=|y_1|=i$, $|x_2|=|y_2|=j$, and $|x_3|=|y_3|$ such that
(i) $x_2\neq y_2$, (ii) $x_1y_2x_3\in L$, and 
(iii) $y_1x_2y_3\in L$. 
\end{lemma}

The above form of our swapping lemma is similar to Lemma \ref{2nd-swapping-lemma}; however, we can no longer choose a pair $(i,j)$ at our will. Moreover, the cardinality of a subset $S$ must be much larger than that in the case of regular languages. Since the proof of this lemma is more involved than that of Lemma \ref{2nd-swapping-lemma}, we postpone it until the next section.

Meanwhile, we see how to use the swapping lemma for context-free languages. First, it is not difficult to show that $Dup$ does not belong to $\mathrm{CFL}/n$ by applying the lemma directly. 

\begin{example}\label{duplicated-cfl}
The language $Dup$ is not in $\mathrm{CFL}/n$ (and thus not in $\cfl$). Let us assume that $Dup\in\mathrm{CFL}/n$ to lead to a contradiction. 
First, choose an advice function $h$ and a context-free language $L$ such that, for any binary string $x$, $x\in Dup$ iff $\track{x}{h(|x|)}\in L$. Let $m$ be a swapping-lemma constant for $L$. 
Second, choose any sufficiently large even number $n$ satisfying that $2^{n/2}> 2mn^2$. Now, let us define a subset $S$ as $S = \left\{\track{x}{h(n)}\in L\mid |x|=n\right\}$. 
It suffices to satisfy the condition that $|S_{i,u}|\leq |S|/kmn$ for any index $i\in[1,n-j_0]_{\integer}$ and any string $u\in\Sigma^{j_0}$. Since $|S|=2^{n/2}$ and 
$|S_{i,u}|= 2^{n/2-|u|}$ for any string $u\in\Sigma^{\leq n/2}$,  we need to set $k=n/2$ and 
$j_0 = \ceilings{\log_{2}{mn^2}}+1$. 
By the swapping lemma for context-free languages, 
there are two indices  $j\in[j_0,k]_{\integer}$ and $i\in[1,n-j]_{\integer}$ and two strings $x = x_1x_2x_3$ and $y=y_1y_2y_3$ in $S$ with $|x_1|=|y_1|=i$, $|x_2|=|y_2|=j$, and $|x_3|=|y_3|$  
such that (i) $x_2\neq y_2$, (ii) $x_1y_2x_3\in L$, 
and (iii) $y_1x_2y_3\in L$.
There are three cases to consider: (a) $i+j\leq n/2$, (b) $i<n/2<i+j$, and (c) $n/2\leq i$. Let us consider Case (a). 
Since $i\geq1$ and $2\leq j\leq n/2$, both $x_2$ and $y_2$ are respectively in the first half portion of $x$ and $y$. Therefore, it is obvious that the swapped strings $x_1y_2x_3$ and $y_1x_2y_3$ are not of the form $\track{ww}{h(n)}$. 
This is a contradiction. The other cases are similar. Therefore, $Dup$ cannot be in $\mathrm{CFL}/n$. 
\qed
\end{example}

In the above example, the choice of $k$ is crucial. For instance, when $k=n/2+1$, there is a case where we cannot lead to any contradiction. Consider the following case. Take two strings $x= x_1x_2x_3$ and $y=y_1y_2y_3$ satisfying that $x_1 = y_1 =0^{n/4-1}$, $x_3=y_3=1^{n/4}$, and 
$x_2=0x_3x_10$, and $y_2=1y_3y_11$. Clearly, $x$ and $y$ are in $L\cap\{0,1\}^{n}$
and the swapped strings $x_1y_2x_3$ and 
$y_1x_2y_3$ are also in $L$.

Our next example is a slightly artificial language $Equal_{6}$, which consists of all strings $w$ over the alphabet $\Sigma=\{a_1,a_2,\ldots,a_{6},\#\}$ such that each symbol except $\#$ in $\Sigma$ appears the same number of times in $w$; that is, $\#_{a}(w)= \#_{b}(w)$ for any pair $a,b\in\Sigma-\{\#\}$.  
Note that the complement $\overline{Equal_{6}}$ 
is in $\mathrm{CFL}$. 
This containment is shown by considering an npda that behaves as follows: on input $w$, choose two distinct symbols, say $a$ and $b$, in $\Sigma-\{\#\}$ nondeterministically and check if $\#_{a}(w) \neq \#_{b}(w)$. In other term, $Equal_{6}$ is in $\co\mathrm{CFL}$. 
On the contrary, we can show that $Equal_{6}$ cannot belong to  $\mathrm{CFL}/n$.

\begin{figure}[ht]
\begin{center}
\includegraphics*[width=13.5cm]{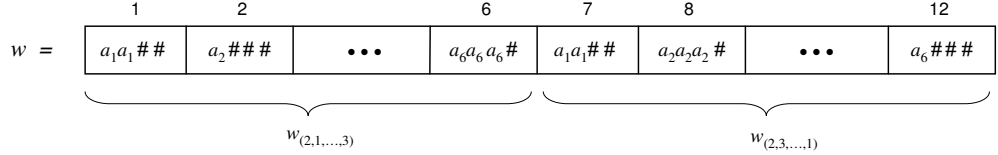}
\caption{A form of $w$ with  $\#_{a_i}(w)=4$ and $\#_{\#}(w)=24$ when $n=48$}\label{fig:Equal6-string} 
\end{center}
\end{figure}

\begin{example}\label{3eqaul-CFL}
The language $Equal_{6}$ is not in $\mathrm{CFL}/n$. 
Assuming that $Equal_{6}\in\mathrm{CFL}/n$, we choose an advice function $h$ and a language $L\in\cfl$ such that, for every string $x\in\Sigma^*$, 
$x\in Equal_{6}$ iff $\track{x}{h(|x|)}\in L$. Since $L\in\cfl$, take a swapping-lemma 
constant $m$ for $L$. 
Let $n  = 864 m$. For each symbol $a_{i}$ in $\Sigma$, we use the notation $a_{(i,e)}$ for the string $(a_{i})^{e}\#^{n/12-e}$ of length $n/12$. 
As a special case, we have $a_{(i,0)} = \#^{n/12}$ and $a_{(i,n/12)} = (a_i)^{n/12}$. For convenience, we denote by $w_{(e_1,e_2,\ldots,e_{6})}$  
the string made up by the following $6$ blocks:  
$a_{(1,e_1)}a_{(2,e_2)}\cdots a_{(6,e_{6})}$. 
Let $S$ be the set consisting of all strings $\track{w}{h(n)}$, where $w$ is  of the form $w_{(e_1,\ldots,e_{6})}w_{(n/12-e_1,\ldots,n/12-e_{6})}$ for any
six indices $e_1,e_2,\ldots,e_{6}\in [0,n/12]_{\integer}$. 
An example of such $w$ is shown in Figure \ref{fig:Equal6-string}.  
Notice that, for any symbol $a\in \Sigma-\{\#\}$, if $w\in S$ then $\#_{a}(w) = n/12$. Moreover, $\#_{\#}(w) = n/2$. Now, we choose $j_0=n/4$ and $k=n/2$. A simple observation gives $|S| = \left(n/12+1\right)^6$. Let $u$ be an arbitrary string in $\Sigma^{j_0}$. 
To estimate $|S_{i,u}|$, we note that $|S_{i,u}| \leq |S_{1,\#^{n/4}}|$  
for any index $i\in[1,n-j_0]_{\integer}$. This gives a simple upper bound: $|S_{i,u}| \leq \left(n/12+1\right)^3$. Obviously, since $n=864m$, we have 
\[
|S_{i,u}| \cdot kmn \leq \left(\frac{n}{12}+1\right)^3 
\cdot \frac{mn^2}{2} 
 =  \frac{m(n+12)^5}{3456} <   \left(\frac{n}{12}+1\right)^6 = |S|.
\] 
The swapping lemma provides an index pair $i,j$ with $n/4 \leq j \leq n/2$ and $i+j\leq n$ and a string pair $x,y\in\Sigma^n$ with $midd_{i,i+j}(x)\neq midd_{i,i+j}(y)$ such that $\track{x}{h(n)},\track{y}{h(n)}\in S$ and $\track{x'}{h(n)},\track{y'}{h(n)}\in L$ , where $x'$ and $y'$ are two swapped strings defined as 
$x' = pref_{i}(x)midd_{i,i+j}(y)suf_{n-i-j}(x)$ and $y' = pref_{i}(y)midd_{i,i+j}(x)suf_{n-i-j}(y)$.  
\sloppy Since the substrings $midd_{i,i+j}(x)$ and $midd_{i,i+j}(y)$ have length $j$, $midd_{i,i+j}(x)\neq midd_{i,i+j}(y)$ implies that $\#_{a}(midd_{i,i+j}(x)) \neq \#_{a}(midd_{i,i+j}(y))$ for a certain symbol $a\in\Sigma-\{\#\}$. Hence, we conclude that $\#_{a}(x')\neq n/12$. This is a contradiction against the statement that $\track{x'}{h(n)}\in L$. 
Therefore, $Equal_{6}$ cannot be in $\mathrm{CFL}/n$.  
\qed
\end{example} 

Since the language $Equal_{6}$ belongs to $\co\mathrm{CFL}$, as shown in Example \ref{3eqaul-CFL}, we can
derive the following strong separation between $\mathrm{CFL}/n$ and $\co\mathrm{CFL}/n$.
 
\begin{proposition}
$\co\mathrm{CFL}\nsubseteq \mathrm{CFL}/n$. Or equivalently,  $\co\mathrm{CFL}/n\neq \mathrm{CFL}/n$.
\end{proposition}

The second part of the above proposition follows from the fact that 
$\co\cfl\subseteq \co\cfl/n$.
This proposition also indicates that $\mathrm{CFL}/n$ is not closed under complementation.

\section{Proof of Lemma \ref{swapping-lemma-CFL}}

As we have seen in Examples \ref{duplicated-cfl} and \ref{3eqaul-CFL}, the swapping lemma for context-free languages are a powerful tool in proving non-context-freeness with advice. 
This section will describe in detail the proof of Lemma \ref{swapping-lemma-CFL}. The proof requires an analysis of a stack's  behavior of a nondeterministic pushdown automaton (or an npda). 

\subsection{Nondeterministic Pushdown Automata}\label{sec:npda}

Although there are several models to describe context-free languages, here, we use the machine model of npda's. We first review certain facts regarding the npda's. 
Let $L$ be any infinite context-free language over an alphabet $\Sigma$ with $|\Sigma|\geq2$. Since Lemma \ref{swapping-lemma-CFL} targets only inputs of length at least $2$,  
it is harmless for us to assume that $L$ contains no empty string $\lambda$. Now, consider a
context-free grammar $G=(V,T,S,P)$ that generates $L$ with $T= \Sigma$,
where $V$ is a set of variables, $T$ is a set of terminal symbols, $S\in V$ is the start variable, and $P$ is a set of productions. Without loss of generality, we can assume that $G$ is in {\em Greibach normal form}; that is, $P$ consists of the production rules of the form $A\,\rightarrow au$, where $A\in V$, $a\in\Sigma$, and $u\in V^*$.
A process of transforming a context-free grammar into Greibach normal form is described in many  undergraduate textbooks (\eg \cite{HU79}).

Closely associated with the grammar $G$, we want to build an npda $M =(Q,\Sigma,\Gamma,\delta,q_0,z,F)$, where $Q$ is a set of internal states, $\Gamma$ is a stack alphabet, $\delta$ is a transition function, $q_0\in Q$ is the initial state, $z\in \Gamma$ is the stack start symbol, and $F\subseteq Q$ is a set of final states. For our npda $M$, we define $Q=\{q_0,q_1,q_f\}$, $z\not\in V$, and $\Gamma = V\cup\{z\}$, $F=\{q_f\}$. 
To make our later argument simpler, we include two special {\em end-markers} 
$\cent$ and $\dollar$, which mark the left end and the right end of an input, respectively. Hereafter, we consider only inputs of the form $\cent x \dollar$, where $x\in\Sigma^*$, and we sometimes treat the endmarkers as an integrated part of the input. Notice that $|\cent x \dollar|=|x|+2$. For convenience, every tape cell is indexed with integers and the left endmarker $\cent$ is always written in the $0$th cell. The input string $x$ of length $n$ is written in the cells indexed between $1$ and $n$ and the right endmarker $\dollar$ is written in the $n+1$st cell. 

When we express the content of the stack of $M$ as a series $s = s_1s_2s_3\cdots s_m$, we understand that the leftmost symbol $s_1$ is located at the top of the stack and the $s_m$ is at the bottom of the stack. At last, the transition function $\delta$ is defined as follows: 
\begin{enumerate}\vs{-1}
\item[(1)] $\delta(q_0,\cent,z) = \{ (q_1,Sz) \}$; 
\vs{-2}
\item[(2)] $\delta(q_1,a,A) = \{ (q_1,u) \mid u\in V^{*}, \text{$P$ contains } A\rightarrow au \}$ for every $a\in\Sigma$ and $A\in V$; and 
\vs{-2}
\item[(3)] $\delta(q_1,\dollar,z) = \{(q_f,z)\}$.
\end{enumerate}\vs{-1}
It is important to note that $M$ is always in the internal state $q_1$ while the tape head scans any cell located between $1$ and $n$.  Note also that, during an accepting computation, the stack of the npda $M$ never becomes empty because of the form of production rules in $P$. Therefore, we can demand that $\delta$ should satisfy the  following requirement. 
\begin{enumerate}\vs{-1}
\item[(4)] For any symbol $a\in\Sigma$, $\delta(q_1,a,z) = \setempty$. 
\end{enumerate}\vs{-1}

Additionally, we modify the npda $M$ and force its stack to increase in size by at most two by encoding several consecutive stack symbols (except for $z$) 
into one new stack symbol. 
For instance, provided that the original npda $M$ increases its stack size by at most $3$, we introduce a new stack alphabet $\Gamma'$ consisting of $(v_1)$, $(v_1v_2)$, and $(v_1v_2v_3)$, where $v_1,v_2,v_3\in\Gamma$. A new transition $\delta'$ is defined as follows. Initially, we define $\delta'(q_0,\cent,z')=\{(q_1,S'z')\}$, where $S'=(S)$ and $z'=(z)$. Consider the case where the top of a new stack contains a new stack symbol $(v_1v_2v_3)$, which indicates that the top three stack symbols of the original computation are $v_1v_2v_3$. If $M$ applies a transition of the form $(q_1,w_1w_2w_3) \in\delta(q_1,a,v_1)$, then we instead apply 
$(q_1,(w_1w_2)(w_3v_2v_3)) \in\delta'(q_1,a,(v_1v_2v_3))$. In case of $(q_1,\lambda)\in\delta(q_1,a,v_1)$, we now apply $(q_1,(v_2v_3))\in\delta'(q_1,a,(v_1v_2v_3))$. 
The other cases of $\delta'$ are similarly defined.  
See, \eg \cite{HU79} for details. 

Overall, we can assume the following extra condition.
\begin{enumerate}\vs{-1}
\item[(5)] for any $a\in\Sigma$, any $v\in\Gamma$, and any $w\in\Gamma^*$, if $(q_1,w)\in \delta(q_1,a,v)$, then $|w|\leq 2$.
\end{enumerate}\vs{-1}

The aforementioned five conditions significantly simplify the proof of Lemma \ref{swapping-lemma-CFL}. In the rest of this paper, we assume that our ndpa $M$ satisfies these conditions. For each string $x\in S$, we write $ACC(x)$ for the set of all accepting computation paths of $M$ on the input $x$. Moreover, let $ACC_n = \bigcup_{x\in S}ACC(x)$. 

\subsection{Stack Transitions, Intervals, and Heights}\label{sec:stack-interval}

\begin{figure}[ht]
\begin{center}
\includegraphics*[width=12cm]{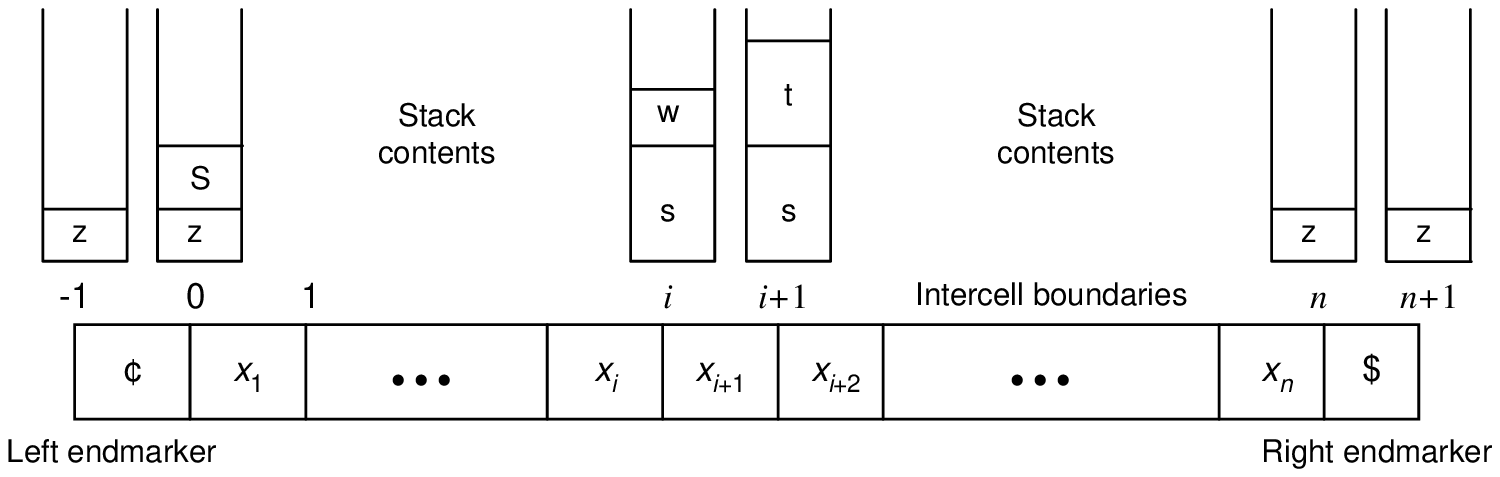}
\caption{An example of intercell boundaries and stack contents}\label{fig:intercell-boundary} 
\end{center}
\end{figure}

For the proof of 
Lemma \ref{swapping-lemma-CFL}, we wish to present our key lemma, Lemma \ref{height-interval}. To describe our lemma, we need to introduce several necessary notions and notations. An {\em intercell boundary} $i$ refers to a boundary or a border between two adjacent cells---the $i$th cell and the $i+1$st cell---in our npda's input tape. 
We sometimes call the intercell boundary $-1$ the {\em initial intercell boundary} and the intercell boundary $n+1$ the {\em final intercell boundary}.
Meanwhile, we fix a subset $S\subseteq L\cap\Sigma^n$, a string $x$ in $S$, and a computation path $p$ of $M$ in $ACC(x)$. 
Along this path $p$, we assign to intercell boundary $i$ 
a stack content produced after scanning the $i$th cell and before scanning the $i+1$st cell. For convenience, such a stack content is referred to as the ``stack content at intercell boundary $i$.'' For instance, the stack contents at the initial and final intercell boundaries are both $z$,   independent of the choice of accepting paths.  
Figure \ref{fig:intercell-boundary} illustrates intercell boundaries and a transition of stack contents.

An accepting computation path of the npda generates a length-$(n+2)$ series $(s_{-1},s_0,s_1,\ldots,s_n,s_{n+1})$ of stack contents with $s_{-1}= s_n = s_{n+1} =z$ and $s_0=Sz$. We refer to this series as a {\em stack transition} associated with the interval $I_0=[-1,n+1]_{\integer}$. More generally, when $I=[i_0,i_1]_{\integer}$ is a subinterval of $I_0$, we call an associated subsequence $\gamma = (s_{i_0},s_{i_0+1},\ldots,s_{i_1})$ 
a {\em stack transition} with the interval $I$. We define the {\em height} at intercell boundary $b$ of $\gamma$ to be the length $|s_{b}|$ of the stack content $s_b$ at $b$. By our choice of the ndpa $M$ given in Section \ref{sec:npda}, 
the minimal height is $1$. 

For our purpose, we hereafter focus our attention only on stack transitions $\gamma$ with intervals $[i_0,i_1]_{\integer}$ in which (i) we have the same height $\ell$ at both of the intercell boundaries $i_0$ and $i_1$ and (ii) all heights within this interval are more than or equal to $\ell$. We briefly call such $\gamma$ {\em ideal}.  

Let $I=[i_0,i_1]_{\integer}$ be any subinterval of $I_0$ and let $\gamma= (s_{i_0},s_{i_0+1},\ldots,s_{i_1})$ be any ideal stack transition with this interval $I$. For each possible height $\ell$, we define the {\em minimal width}, denoted $minwid_{I}(\ell)$ (resp., the {\em maximal width}, denoted $maxwid_{I}(\ell)$), to be the minimal value (resp., maximal value) $|I'|=i'_1-i'_0$ for which (i) $I'=[i'_0,i'_1]_{\integer}\subseteq I$, (ii) $\gamma$ has height $\ell$ at both intercell boundaries $i'_0$ and $i'_1$, and (iii) at no intercell boundary $i\in I'$, $\gamma$ has height less than $\ell$. Such a pair $(i'_0,i'_1)$ produces a subsequence $\gamma'=(s_{i'_0},s_{i'_0+1},\ldots,s_{i'_1})$ of $\gamma$. In such a case, we say that $\gamma'$ {\em realizes} the minimal width $minwid_{I}(\ell)$ (resp., maximal width $maxwid_{I}(\ell)$).  

\begin{figure}[ht]
\begin{center}
\includegraphics*[width=11.0cm]{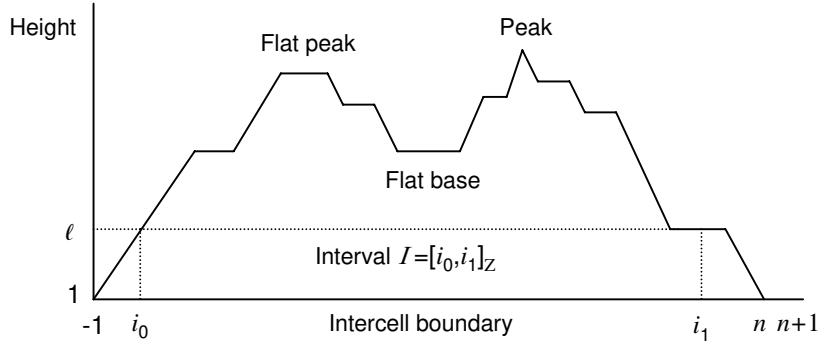}
\caption{An example of track transition with interval $I=[i_0,i_1]_{\integer}$ and height $\ell$}\label{fig:stack-transition} 
\end{center}
\end{figure}

We say that a stack transition $\gamma$ has a {\em peak at $i$} if $|s_{i-1}|<|s_i|$ and $|s_{i+1}|<|s_i|$. Moreover, $\gamma$ has a {\em flat peak in $(i'_0,i'_1)$} if $|s_{i'_0-1}|<|s_{i'_0}|=|s_{i'_0+1}|=\cdots = |s_{i'_1}|$ and $|s_{i'_1+1}|<|s_{i'_1}|$. 
On the contrary, we say that $\gamma$ has a {\em base at $i$} if 
$|s_{i-1}|>|s_i|$ and $|s_{i+1}|>|s_i|$; $\gamma$ has a {\em flat base in $(i'_0,i'_1)$} if $|s_{i'_0-1}|>|s_{i'_0}|=|s_{i'_0+1}|=\cdots = |s_{i'_1}|$ and $|s_{i'_1+1}|>|s_{i'_1}|$. See Figure \ref{fig:stack-transition} for an example of (flat) peaks and (flat) bases.

At last, we state our key lemma, which holds for any accepting computation path $p$ without any assumption other than $j_0\geq2$ and $2j_0\leq k \leq n$. 

\begin{lemma}{\rm [key lemma]}\label{height-interval}
Let $S\subseteq L\cap\Sigma^n$, let $x$ be any string in $S$, and let $p$ be any computation path of $M$ in $ACC(x)$. Assume that $j_0\geq2$ and $2j_0\leq k \leq n$. For any interval $I =[i_0,i_1]_{\integer} \subseteq[-1,n+1]_{\integer}$ with $|I|>k$ and for any ideal stack transition $\gamma$ with the interval $I$ having height $\ell_0$ at the two 
intercell boundaries $i_0$ and $i_1$, there are a subinterval $I'=[i'_0,i'_1]_{\integer}$ of $I$ and a height $\ell\in[1,n]_{\integer}$ 
such that $\gamma$ has height $\ell$ at both intercell boundaries $i'_0$ and $i'_1$, $j_0\leq |I'|\leq k$, and $minwid_{I}(\ell)\leq |I'| \leq maxwid_{I}(\ell)$.
\end{lemma}

\begin{proof}
Fix six parameters $(x,p,\gamma,\ell_0,I)$ given in the premise of the lemma. We prove the lemma by induction on the number of peaks or flat peaks along the computation path $p$ of $M$ in $ACC(x)$. 

\s

(Basis Case) In this particular case, there is either one peak or one flat peak 
in $\gamma$ in the interval $I=[i_0,i_1]_{\integer}$. First, we consider the case where there is a peak. 
Let $\ell_1$ be the height of such a peak. 
Note that $minwid_{I}(\ell) = maxwid_{I}(\ell+1) +2$ for any height $\ell$ with $\ell_0\leq \ell<\ell_1$, because of the condition 5 
provided for the npda's transition function $\delta$. 
Now, let $\ell'$ be the maximal height satisfying that $minwid_{I}(\ell'+1)\leq j_0 < minwid_{I}(\ell')$. 
Such $\ell'$ exists because $|I|>k > j_0$.
Let $I_{min}=[i'_0,i'_1]_{\integer}$ be the {\em minimal} interval such that $\gamma$ has height $\ell'+1$ at the two 
intercell boundaries $i'_0$ and $i'_1$. Similarly, let $I_{max} = [i''_0,i''_1]_{\integer}$ be the {\em maximal} interval that satisfies a similar condition with $\ell'+1$, $i''_0$, and $i''_1$. 
If $j_0= minwid_{I}(\ell'+1)$, then we choose  
the desired interval $I'=I_{min}$ and the height $\ell=\ell'+1$ for the lemma. 
If $j_0\leq maxwid_{I}(\ell'+1)$, then we pick  
an interval $I'$ satisfying that $I_{min}\subseteq I'\subseteq I_{max}$ and $|I'|=j_0$. We also define $\ell=\ell'+1$ for the lemma.
The remaining case to consider is that $maxwid_{I}(\ell'+1) < j_0 < minwid_{I}(\ell')$. In this case, it follows that 
\[
j_0< minwid_{I}(\ell') = maxwid_{I}(\ell'+1)+2 < j_0 + 2 \leq 2j_0 \leq k
\]
since $j_0\geq 2$. Let $I'_{min}=[\hat{i}_0,\hat{i}_1]_{\integer}$ be the minimal interval such that $\gamma$ has height $\ell'$ at the two intercell boundaries $\hat{i}_0$ and $\hat{i}_1$. It is thus enough to define $I'=I'_{min}$ and $\ell= \ell'$ for the lemma.   

Next, we consider the case where there is a flat peak in $(i_2,i_3)$ with height $\ell_1$. 
If $i_3-i_2 \geq j_0$, then we choose $I'=[i_2,i_2+j_0]_{\integer}$ and $\ell=\ell_1$ for the lemma. The other case where $i_3-i_2 < j_0$ is similar to the ``peak'' case discussed above.

\s

(Induction Step) First, let $c>1$ and consider the case where there are $c$ peaks and/or flat peaks in the given interval $I=[i_0,i_1]_{\integer}$ with $|I|>k$. Choose the lowest base or flat base within this interval. 
If we have more than one such base and/or flat base, then we always choose the leftmost one. Now, consider the case where 
there is the lowest base at $i_2$ and let $\ell_2$ be the height at $i_2$. Since $\gamma$ is an ideal stack transition, we have $\ell_2\geq \ell_0$. Let  $I^*=[i'_0,i'_1]_{\integer}$ be the {\em largest} interval for which the height at both $i'_0$ and $i'_1$ equals $\ell_2$. The choice of $I^*$ implies that $|I^*| = maxwid_{I}(\ell_2)$. We split $I^*$ into two subintervals $I_1=[i'_0,i_2]_{\integer}$ and $I_2 = [i_2,i'_1]_{\integer}$.  
If $j_0\leq |I^*|\leq k$, then we set $I'=I^*$ and $\ell=\ell_2$ for the lemma. 
If $|I^*|<j_0$, then a similar argument used for the basis case proves the lemma. 
Now, assume that $|I^*|> k$. Since $k\geq 2j_0$, either one of of $I_1$ and $I_2$ has size more than $j_0$. We pick such an interval, say $I_3$.  Let $\gamma'$ be a unique subsequence of $\gamma$ defined in the interval $I_3$.
If $|I_3|\leq k$, 
then we choose $I'=I_3$ and $\ell=\ell_2$ for the lemma. 
Let us assume that $|I_3|>k$. By the choice of $I_3$, 
$\gamma'$ is an ideal stack transition. Since $\gamma'$ has fewer than $c$ peaks and/or flat peaks, we can apply the induction hypothesis to obtain the lemma. 

Next, we consider the other case where there is the lowest flat base in $(i_2,i_3)$. We define $I^*=[i'_0,i'_1]_{\integer}$ as before. Unlike the previous ``lowest base'' case, we need to split $I^*$ into three intervals $I_1=[i'_0,i_2]_{\integer}$, $I_2=[i_2,i_3]_{\integer}$, and $I_3=[i_3,i'_1]_{\integer}$. 
If either $|I^*|<j_0$ or $j_0\leq |I^*|\leq k$, then it suffices to apply a similar argument used for the ``lowest base'' case. Now, assume that $|I^*|>k$. Since $k\geq 2j_0$, either one of the two intervals $I_1\cup I_2$  and $I_3$ has size more than $j_0$. 
We pick such an interval. The rest of our argument is similar to the one for the ``lowest base'' case.
\end{proof}

\subsection{Technical Tools}

Let $M=(Q,\Sigma,\Gamma,\delta,q_0,z,F)$ be our npda for $L$ defined in Section \ref{sec:npda}. We have already seen fundamental properties of our npda in Section \ref{sec:stack-interval}. Now, let us begin proving Lemma \ref{swapping-lemma-CFL} by contradiction. 
First, we set our swapping-lemma constant $m$ to be $|\Gamma|^2$ and 
assume that the conclusion of Lemma \ref{swapping-lemma-CFL} is false for this $m$; 
that is, the following assumption (a) holds for four fixed parameters $(n,j_0,k,S)$ given in the premise of the lemma. We fix these parameters throughout this subsection and its subsequent subsection. 
\begin{itemize}\vs{-1}
\item[(a)] There are no indices $i\in[1,n]_{\integer}$ and $j\in[j_0,k]_{\integer}$ with $i+j\leq n$ and no strings $x =x_1x_2x_3$ and $y=y_1y_2y_3$ in $S$ with $|x_1|=|y_1|=i$, $|x_2|=|y_2|=j$, and $|x_3|=|y_3|$ such that 
(i) $x_2\neq y_2$, (ii) $x_1y_2x_3\in L$, and 
(iii) $y_1x_2y_3\in L$. 
\end{itemize}

Recall that $S$ is a fixed subset of $L\cap\Sigma^n$. Meanwhile, we fix additional five 
parameters $x\in S$,  $j\in[j_0,k]_{\integer}$, $i\in[1,n-j]_{\integer}$, $v\in\Gamma$, and $p\in ACC(x)$.
As a technical tool, we introduce the notation $G_{i,j,p}(x:v)$. 
Roughly speaking, $G_{i,j,p}(x:v)$ expresses a part of stack content that is newly produced from its original content $vs$ during the head's scanning the cells indexed between $i+1$ and $i+j$, provided that the npda scans no symbol in $s$. Note that, when the npda is deterministic, the information on $p$ can be discarded, because $p$ is completely determined 
from $x$. More precisely, $G_{i,j,p}(x:v)$ denotes
a unique string $t\in\Gamma^*$ (if any) that satisfies the following three conditions, along the computation path $p$ with the input $x$. 
\begin{enumerate}\vs{-1}
\item The stack consists of $vs$ at the intercell boundary $i$, where $s\in \Gamma^*$.
\vs{-2}
\item At the intercell boundary $i+j$, the stack consists of $ts$.
\vs{-2}
\item While the head scans any cell indexed between $i+1$ and $i+j$, the npda never accesses any symbol in $s$; that is, no transition of the form $(q_1,w)\in \delta(q_1,a,r)$, where $r$ is a symbol in $s$, is applied.
\end{enumerate}\vs{-1}

With the fixed parameters $(i,j,v,t,p)$ described above, 
we use the notation 
$T^{(i)}_{j,v,t,p}$ to denote the set $\{x\in S\mid G_{i,j,p}(x:v)=t\}$. A crucial property of $T^{(i)}_{j,v,t,p}$ is stated in the following lemma.

\begin{lemma}\label{change-middle}
We fix $j\in[j_0,k]_{\integer}$, $i\in[1,n-j]_{\integer}$, $p,p'\in ACC_n$, $v\in \Gamma$, and $t\in\Gamma^*$. For any two strings $x,y$ in $S$, if $x\in T^{(i)}_{j,v,t,p}$ and $y\in T^{(i)}_{j,v,t,p'}$, then two swapped strings $pref_{i}(x)midd_{i,i+j}(y)suf_{n-i-j}(x)$ and $pref_{i}(y)midd_{i,i+j}(x)suf_{n-i-j}(y)$ are both in $L$. 
\end{lemma}

\begin{proof}
Assume that the npda's stack consists of $vs$ (resp., $vs'$) at an intercell boundary $i$ along an accepting path $p$ (resp., $p'$) on an input $x$ (resp., $y$). Since $x\in T^{(i)}_{j,v,t,p}$ (resp.,  $y\in T^{(i)}_{j,v,t,p'}$), the npda generates a stack content $ts$ 
(resp., $ts'$) at the intercell boundary $i+j$. 
Note that, during the head's scanning the cells indexed between 
$i+1$ and $i+j$, the npda never accesses $s$ (resp., $s'$) along the path $p$ (resp., $p'$). 
Therefore, we
can swap two substrings  $midd_{i,i+j}(x)$ and $midd_{i,i+j}(y)$  written in the cells indexed between 
$i+1$ and $i+j$ in $x$ and $y$, respectively, without changing the acceptance condition of the npda. Therefore, the npda accepts the two strings $pref_{i}(x)midd_{i,i+j}(y)suf_{n-i-j}(x)$ and $pref_{i}(y)midd_{i,i+j}(x)suf_{n-i-j}(y)$. This implies the conclusion of the lemma.
\end{proof}

Recall from Section \ref{sec:CFL} the notation $S_{i,w}$. Now, we consider the following statement.
\begin{itemize}\vs{-1}
\item[(b)] There are no $j\in[j_0,k]_{\integer}$, $i\in[1,n-j]_{\integer}$, $p,p'\in ACC_n$,  $t\in\Gamma^*$, or $u,w\in\Sigma^{j}$ such that $u\neq w$, $T^{(i)}_{j,v,t,p}\cap S_{i,u} \neq\setempty$, and $T^{(i)}_{j,v,t,p'}\cap S_{i,w} \neq\setempty$. 
\end{itemize}\vs{-1}

\begin{lemma}\label{statement-b}
The statement (a) implies the statement (b). 
\end{lemma}

\begin{proof}
Assume the statement (a). To show the statement (b), 
let us assume on the contrary 
that the statement (b) does not hold. 
This means that certain parameters $(i,j,p,p',t,u,w)$ satisfy 
the following conditions:  
$u\neq w$, $T^{(i)}_{j,v,t,p}\cap S_{i,u} \neq\setempty$, and $T^{(i)}_{j,v,t,p'}\cap S_{i,w} \neq\setempty$.
Now, we take two strings $x\in T^{(i)}_{j,v,w,p}\cap S_{i,u}$ and $y\in T^{(i)}_{j,v,w,p'}\cap S_{i,w}$. Notice that $|u|=|w|=j$. Since $x\in S_{i,u}$ and $y\in S_{i,w}$, it follows that $u = midd_{i,i+j}(x)$ and $w = midd_{i,i+j}(y)$. Lemma \ref{change-middle} then implies that the swapped strings $x' = pref_{i}(x)midd_{i,i+j}(y) suf_{n-i-j}(x)$ and $y' = pref_{i}(y) midd_{i,i+j}(x) suf_{n-i-j}(y)$ are both in $L$. This contradicts the statement (a). Therefore, the statement (b) holds. 
\end{proof}

{}From Lemmas \ref{change-middle} and \ref{statement-b}, the choice of accepting paths for strings in $S$ is of little importance. Hence, for later convenience, we write $T^{(i)}_{j,v,t}$ to denote the union $\bigcup_{p\in ACC_n} T^{(i)}_{j,v,t,p}$.

\subsection{Closing Argument}

Under our assumption that Lemma \ref{swapping-lemma-CFL} is false, 
we want to lead to a desired contradiction, 
which immediately proves the lemma. To achieve our goal, we utilize Lemma \ref{height-interval} given in Section \ref{sec:stack-interval}. 
Notice that, by Lemma \ref{statement-b}, 
the statement (b) now holds. Recall also that three parameters $(n,j_0,k,S)$ are fixed through our proof.

For our convenience, we write $\Delta$ for the index set $\{(i,j,v,w)\mid j\in[j_0,k]_{\integer},i\in[1,n-j]_{\integer},v,w\in\Gamma\}$. The cardinality of this set $\Delta$ is equal to 
\[
|\Delta| = (k-j_0)(n-j_0+1)|\Gamma|^2 = m(k-j_0+1)(n-j_0+1)
\]
since $m=|\Gamma|^2$. 
We want to assign each string $x$ in $S$ to a certain element $(i,j,v,w)$ in $\Delta$. For this purpose, we first show the following lemma, which can be obtained immediately from Lemma \ref{height-interval}. 

\begin{lemma}\label{good-interval}
Assume that the statement (a) holds. 
Let $j_0$ and $k$ satisfy that $j_0\geq 2$ and $2j_0 \leq k\leq n$. 
For any string $x\in S$, there is an element $(i,j,v,w)\in \Delta$ for which  $x\in T^{(i)}_{j,v,w}$.
\end{lemma}

\begin{proof}
Let $x$ be any string in $S$ and consider any computation path $p$ of the npda $M$ in $ACC(x)$. 
We denote by $\gamma$ the stack transition of $M$ with the interval $I_0=[-1,n+1]_{\integer}$. 
Lemma \ref{height-interval} guarantees, in the stack transition $\gamma$, the existence of a subinterval $I'=[i'_0,i'_1]_{\integer}$ and a height $\ell$  satisfying that $j_0\leq |I'| \leq k$, $minwid_{I}(\ell)\leq |I'| \leq maxwid_{I}(\ell)$, and $\gamma$ has height $\ell$ at both of the intercell boundaries $i'_0$ and $i'_1$. 
For our desired index values $i$ and $j$, 
we define them as $i=i'_0$ and $j = |I'|$ ($=i'_1-i'_0$). 
Now, let us consider the changes of stack contents while $M$'s head scans through the interval $I'$. Let $vs$ be the stack content at the intercell boundary $i'_0$ and let $ws'$ be the stack content at $i'_1$, 
where $v,w\in\Gamma$ and $s,s'\in\Gamma^*$. Note that $\ell=|vs|=|ws'|$. Since all heights in $I'$ are at least $\ell$, within this interval $I'$, $M$ does not access any symbol in $s$;  hence, we conclude that $s=s'$. This implies that $G_{i,j,p}(x:v)$ equals $w$. Therefore, $x$ should be in $T^{(i)}_{j,v,w}$.
\end{proof}

Lemma \ref{good-interval} establishes a key association of strings in $S$ with elements in $\Delta$. Using this association, we introduce a map $e$ from $S$ to $\Delta$. For each string $x$ in $S$, assuming an appropriate lexicographic order for $\Delta$, we denote by $e(x)$ the {\em minimal} element $(i,j,v,w)\in \Delta$ satisfying that $x\in T^{(i)}_{j,v,w}$. 
Notice that this minimality requirement 
makes $e(x)$ uniquely determined from $x$. 
With this map $e$, we define $A_{i,j,v,w}$ as the set 
$\{x\in S\mid e(x) = (i,j,v,w)\}$. Obviously, it follows that  $A_{i,j,v,w}\subseteq T^{(i)}_{j,v,w}$.

Now, we claim the following property of the map $e$.

\begin{claim}\label{encoding}
There exist two strings $x,y\in S$ and also 
two strings $u,z\in\Sigma^{j}$ such that $u\neq z$, 
$x\in S_{i,u}$,  $y\in S_{i,z}$, and $e(x)=e(y)=(i,j,v,t)$. 
\end{claim}

If this claim is true, then we take four strings $(x,y,u,z)$ given in the claim. Since $e(x)=e(y)=(i,j,w,t)$, we obtain $x,y\in A_{i,j,v,w}$.  
Since $A_{i,j,v,w}\subseteq T^{(i)}_{j,v,w}$, it follows that
$x\in T^{(i)}_{j,v,w}\cap S_{i,u}$ and $y\in T^{(i)}_{j,v,w}\cap S_{i,z}$. 
This is obviously a contradiction against the assumption (b) 
and hence the assumption (a). 
Therefore, Lemma \ref{swapping-lemma-CFL} should hold. 

What remains undone is the verification of Claim \ref{encoding}. Let us prove the claim. 
Since $e(\cdot)$ is a map from $S$ to $\Delta$, there is a certain element $(i,j,v,w)\in\Delta$ satisfying that $|A_{i,j,v,w}|\geq |S|/|\Delta|$. 
Fix such an element $(i,j,v,w)$. 
For any string $u\in\Sigma^{j}$, since $|\Delta| = m(k-j_0+1)(n-j_0+1)$, 
we obtain  
\[
|S_{i,u}|< \frac{|S|}{m(k-j_0+1)(n-j_0+1)} = \frac{|S|}{|\Delta|} \leq |A_{i,j,v,w}|,
\]
where the first inequality is one of the premises of Lemma \ref{swapping-lemma-CFL}.
Since $S=\bigcup_{u\in\Sigma^j} S_{i,u}$, the above inequality concludes that there are at least two distinct strings $u,z\in\Sigma^j$ for which certain strings  $x\in S_{i,u}$ and $y\in S_{i,z}$ map to the same element  $(i,j,v,t)$. This completes the proof of the claim and therefore  completes the proof of Lemma \ref{swapping-lemma-CFL}.

\bs\bs
\n{\bf Acknowledgments.} 
The author thanks Satoshi Okawa and Francois Le Gall for a helpful 
discussion on a fundamental structure of context-free languages. 

\bibliographystyle{alpha}

\end{document}